\def\fracm#1#2{\hbox{\large{${\frac{{#1}}{{#2}}}$}}}

\documentstyle[12pt]{article}

\def\@magscale#1{ scaled \magstep #1}

\catcode`@=11  
\def\un#1{\relax\ifmmode\@@underline#1\else
        $\@@underline{\hbox{#1}}$\relax\fi}
\catcode`@=12

\def\a{\alpha}
\def\b{\beta}

\def\d{\delta}
\def\e{\epsilon}

\def\g{\gamma}

\def\l{\lambda}

\def\o{\omega}
\def\p{\pi}
\def\q{\theta}

\def\D{\Delta}

\def\G{\Gamma}

\def\L{\Lambda}

\def\dslash{\not{\hbox{\kern-2pt $\partial$}}}
\def\Dslash{\not{\hbox{\kern-4pt $D$}}}
\def\pslash{\not{\hbox{\kern-2.3pt $p$}}}
 \newtoks\slashfraction
 \slashfraction={.13}
 \def\slash#1{\setbox0\hbox{$ #1 $}
 \setbox0\hbox to \the\slashfraction\wd0{\hss \box0}/\box0 }
 
\font\ro=cmsy10                          % font with rope
\def\kcr{{\hbox{\ro \char'170}}}                % right-handed rope
\def\ktl{{\hbox{\ro \char'170}}}        % top end for left-handed rope
\def\ktr{{\hbox{\ro \char'170}}}        % " right
\def\kbl{{\hbox{\ro \char'170}}}        % " bottom left
\def\kbr{{\hbox{\ro \char'170}}}        % " right
\def\plpl{\raise-2pt\hbox{$\raise3pt\hbox{$_+$}\hskip-6.67pt\raise0.0pt
\hbox{$^+$}\hskip 0.01pt$}}
\def\mimi{\raise-2pt\hbox{$\raise3pt\hbox{$_-$}\hskip-6.67pt\raise0.0pt
\hbox{$^-$}\hskip 0.01pt$}} 

\def\bo{{\raise.15ex\hbox{\large$\Box$}}}               % D'Alembertian
\def\pa{\partial}                                       % curly d
\def\TH{{\raise.2ex\hbox{$\displaystyle \bigodot$}\mskip-4.7mu \llap H
\;}}
\def\face{{\raise.2ex\hbox{$\displaystyle \bigodot$}\mskip-2.2mu \llap
{$\ddot
        \smile$}}}                                      % happy face
\def\sp#1{{}^{#1}}                              % superscript (unaligned)
\def\Tilde#1{\widetilde{#1}}                    % big tilde
\def\Hat#1{\widehat{#1}}                        % big hat
\def\Bar#1{\overline{#1}}                       % big bar
\def\leftrightarrowfill{$\mathsurround=0pt \mathord\leftarrow \mkern-6mu
        \cleaders\hbox{$\mkern-2mu \mathord- \mkern-2mu$}\hfill
        \mkern-6mu \mathord\rightarrow$}
\def\dvec#1{\vbox{\ialign{##\crcr
        \leftrightarrowfill\crcr\noalign{\kern-1pt\nointerlineskip}
        $\hfil\displaystyle{#1}\hfil$\crcr}}}           % <--> accent
\def\fracm#1#2{\hbox{\large{${\frac{{#1}}{{#2}}}$}}}
\def\frac#1#2{{\textstyle{#1\over\vphantom2\smash{\raise.20ex
        \hbox{$\scriptstyle{#2}$}}}}}                   % fraction
\def\sfrac#1#2{{\vphantom1\smash{\lower.5ex\hbox{\small$#1$}}\over
        \vphantom1\smash{\raise.4ex\hbox{\small$#2$}}}} % alternate fraction
\def\bfrac#1#2{{\vphantom1\smash{\lower.5ex\hbox{$#1$}}\over
        \vphantom1\smash{\raise.3ex\hbox{$#2$}}}}       % "
\def\afrac#1#2{{\vphantom1\smash{\lower.5ex\hbox{$#1$}}\over#2}}    % "
\newskip\humongous \humongous=0pt plus 1000pt minus 1000pt
\def\caja{\mathsurround=0pt}
\def\eqalign#1{\,\vcenter{\openup2\jot \caja
        \ialign{\strut \hfil$\displaystyle{##}$&$
        \displaystyle{{}##}$\hfil\crcr#1\crcr}}\,}
\newif\ifdtup

\def\ref#1{$\sp{#1)}$}

\parskip=\medskipamount                 % space between paragraphs (LaTeX)
\thicklines                         % thick straight lines for pictures (LaTeX)

\def\oldheadpic{                                % old UM heading
        \setlength{\unitlength}{.4mm}
        \thinlines
        \par
        \begin{picture}(349,16)
        \put(325,16){\line(1,0){4}}
        \put(330,16){\line(1,0){4}}
        \put(340,16){\line(1,0){4}}
        \put(335,0){\line(1,0){4}}
        \put(340,0){\line(1,0){4}}
        \put(345,0){\line(1,0){4}}
        \put(329,0){\line(0,1){16}}
        \put(330,0){\line(0,1){16}}
        \put(339,0){\line(0,1){16}}
        \put(340,0){\line(0,1){16}}
        \put(344,0){\line(0,1){16}}
        \put(345,0){\line(0,1){16}}
        \put(329,16){\oval(8,32)[bl]}
        \put(330,16){\oval(8,32)[br]}
        \put(339,0){\oval(8,32)[tl]}
        \put(345,0){\oval(8,32)[tr]}
        \end{picture}
        \par
        \thicklines
        \vskip.2in}
\def\oldtitle#1#2#3#4{\oldheadpic\begin{center}\vglue.5in{\large\bf
#1}\\[.6in]
        {#2}\\[.1in] {\it Department of Physics and Astronomy}\\
        {\it University of Maryland, College Park, MD 20742}\\[.6in]
        Physics Publication \#{#3}\\ {#4}\\[1.5in] {\bf
ABSTRACT}\\[.1in]
        \end{center} \begin{quotation}}                 % old title stuff
\def\oldTitle#1#2#3#4#5#6#7{\oldheadpic\begin{center} \vglue .4in
        {\large\bf #1}\\[.4in]
        {#2}\\[.1in] {\it Department of Physics and Astronomy}\\
        {\it University of Maryland, College Park, MD 20742}\\[.1in]
        {#3}\\[.1in] {\it {#4}}\\ {\it {#5}}\\[.4in]
        Physics Publication \#{#6}\\ {#7}\\[.5in] {\bf ABSTRACT}\\[.1in]
        \end{center} \begin{quotation}}                 % " for 2 authors
\def\border{                                            % border
        \setlength{\unitlength}{1mm}
        \newcount\xco
        \newcount\yco
        \xco=-21
        \yco=12
        \begin{picture}(140,0)
        \put(\xco,\yco){$\ktl$}
        \advance\yco by-1
        {\loop
        \put(\xco,\yco){$\kcr$}
        \advance\yco by-2
        \ifnum\yco>-240
        \repeat
        \put(\xco,\yco){$\kbl$}}
        \xco=158
        \yco=12
        \put(\xco,\yco){$\ktr$}
        \advance\yco by-1
        {\loop
        \put(\xco,\yco){$\kcr$}
        \advance\yco by-2
        \ifnum\yco>-240
        \repeat
        \put(\xco,\yco){$\kbr$}}
        \put(-20,13){\tiny University of Maryland Elementary Particle
Physics University of Maryland Elementary Particle Physics University of
Maryland Elementary Particle Physics}
        \put(-20,-241.5){\tiny University of Maryland Elementary
Particle Physics University of Maryland Elementary Particle Physics
University of Maryland Elementary Particle Physics}
        \end{picture}
        \par\vskip-8mm}
\def\bordero{                                           % alternate border
        \setlength{\unitlength}{1mm}
        \newcount\xco
        \newcount\yco
        \xco=-31
        \yco=12
        \begin{picture}(140,0)
        \put(\xco,\yco){$\ktl$}
        \advance\yco by-1
        {\loop
        \put(\xco,\yco){$\kclr}
        \advance\yco by-2
        \ifnum\yco>-240
        \repeat
        \put(\xco,\yco){$\kbl$}}
        \xco=151
        \yco=12
        \put(\xco,\yco){$\ktr$}
        \advance\yco by-1
        {\loop
        \put(\xco,\yco){$\kcr$}
        \advance\yco by-2
        \ifnum\yco>-240
        \repeat
        \put(\xco,\yco){$\kbr$}}
        \put(-20,12){\ooo
bacdefghidfghghdhededbihdgdfdfhhdheidhdhebaaahjhhdahba

hgdedge
   hgfdiehhgdigicba}
        \put(-20,-241.5){\ooo
ababaighefdbfghgeahgdfgafagihdidihiidhiagfedhadbfd

ecdcdfa
   gdcbhaddhbgfchbgfdacfediacbabab}
        \end{picture}
        \par\vskip-8mm}
\def\headpic{                                           % UM heading
        \indent
        \setlength{\unitlength}{.4mm}
        \thinlines
        \par
        \begin{picture}(29,16)
        \put(165,16){\line(1,0){4}}
        \put(170,16){\line(1,0){4}}
        \put(180,16){\line(1,0){4}}
        \put(175,0){\line(1,0){4}}
        \put(180,0){\line(1,0){4}}
        \put(185,0){\line(1,0){4}}
        \put(169,0){\line(0,1){16}}
        \put(170,0){\line(0,1){16}}
        \put(179,0){\line(0,1){16}}
        \put(180,0){\line(0,1){16}}
        \put(184,0){\line(0,1){16}}
        \put(185,0){\line(0,1){16}}
        \put(169,16){\oval(8,32)[bl]}
        \put(170,16){\oval(8,32)[br]}
        \put(179,0){\oval(8,32)[tl]}
        \put(185,0){\oval(8,32)[tr]}
        \end{picture}
        \par\vskip-6.5mm
        \thicklines}
\def\title#1#2#3#4{\border\headpic {\hbox to\hsize{#4 \hfill UMDEPP
#3}}\par
        \begin{center} \vglue .5in {\large\bf #1}\\[.6in]
        {#2}\\[.1in] {\it Department of Physics and Astronomy}\\
        {\it University of Maryland, College Park, MD 20742}\\[1.5in]
        {\bf ABSTRACT}\\[.1in] \end{center} \begin{quotation}}  
\def\Title#1#2#3#4#5#6#7{\border\headpic
        {\hbox to\hsize{#7 \hfill UMDEPP #6}}\par
        \begin{center} \vglue .4in {\large\bf #1}\\[.4in]
        {#2}\\[.1in] {\it Department of Physics and Astronomy}\\
        {\it University of Maryland, College Park, MD 20742}\\[.1in]
        {#3}\\[.1in] {\it {#4}}\\ {\it {#5}}\\[.5in] {\bf
ABSTRACT}\\[.1in]
        \end{center} \begin{quotation}}                 % " for 2 authors
\def\endtitle{\end{quotation}\newpage}                  % end title page

\def\ad{{\kern0.5pt
                   \alpha \kern-5.05pt
\raise5.8pt\hbox{$\textstyle.$}\kern
0.5pt}}

\def\bd{{\kern0.5pt
                   \beta \kern-5.05pt
\raise5.8pt\hbox{$\textstyle.$}\kern
0.5pt}}

\def\qd{{\kern0.5pt
                   q \kern-5.05pt \raise5.8pt\hbox{$\textstyle.$}\kern
0.5pt}}
\def\Dot#1{{\kern0.5pt
                   {#1} \kern-5.05pt
\raise5.8pt\hbox{$\textstyle.$}\kern
0.5pt}}

\begin{document}

\def\gfrac#1#2{\frac {\scriptstyle{#1}}
        {\mbox{\raisebox{-.6ex}{$\scriptstyle{#2}$}}}}
\def\gg{{\hbox{\sc g}}}
\border\headpic {\hbox to\hsize{February 2000 \hfill
{Bicocca--FT--00--01}}}
\par \hfill {BRX TH-470}
\par \hfill {McGill 00-05}
\par \hfill {UMDEPP 00-049}
\par
\setlength{\oddsidemargin}{0.3in}
\setlength{\evensidemargin}{-0.3in}
\begin{center}
\vglue .1in
{\large\bf Holomorphy, Minimal Homotopy and the\\ 4D, $N$ = 1 
Supersymmetric\\ Bardeen-Gross-Jackiw Anomaly 
Terms\footnote{Supported in part by National Science Foundation 
Grants PHY-98-02551, PHY-9604587  \newline ${~~~~~}$ and 
by MURST Grant 12/2-11/480.}  }
\\[.25in]
S. James Gates, Jr.\footnote{gatess@wam.umd.edu}
\\[0.06in]
{\it Department of Physics, 
University of Maryland\\ College Park, MD 20742-4111 USA}
\\[.1in] Marcus T. Grisaru\footnote{grisaru@brandeis.edu}\footnote{
On leave from Brandeis University}
\\[0.06in]
{\it  Physics Department, McGill University\\ 
Montreal, QC Canada  H3A 2T8}\\[.1in]
and \\ [.1in] Silvia Penati\footnote{Silvia.Penati@mi.infn.it}
\\[0.06in]
{\it Dipartmento di Fisica dell'Universit\' a di Milano--Bicocca \\
and INFN, Sezione di Milano, via Celoria 16, I-20133 Milano,
Italy}\\[.3in]

{\bf ABSTRACT}\\[.01in]
\end{center}
\begin{quotation}
{By use of a special homotopy operator, we present an explicit,
closed-form and {\it {simple}} expression for the left-right 
Bardeen-Gross-Jackiw anomalies described as the proper superspace 
integral of a superfunction. }  

${~~~}$ \newline

\endtitle

\noindent
{\bf {(I.) Introduction}}  

Years ago, Bardeen \cite{A} as well as Gross and Jackiw \cite{B} 
(BGJ) considered the question of the simultaneous quantum consistency 
of conservation laws involving vector and axial vector non-Abelian 
currents.   The results of these studies are now widely appreciated.  
While classically both types of currents can be simultaneously 
conserved, when the effects of relativistic quantum theory are taken 
into account both currents {\it {cannot}} be simultaneously conserved.
One of the conservation laws must be broken due to an anomaly in the  
corresponding Ward identities. The implications of the anomaly are,  
in the words of Zumino, Wu and Zee, ``ubiquitous'' \cite{C}.   A topic
closely related to this is the form of the WZNW term \cite{C1} whose 
variation produces the appropriate anomalies, and its extension in 
the presence of 4D, $N=1$ supersymmetry.

Earlier work on the latter subject has framed the issue by formulating 
a superspace WZNW action described solely in terms of chiral superfields. 
However, in ref. \cite{C2}  a surprising alternative  proposal has
been made.  Namely, if the spin-1/2 fields (``pionini'') accompanying
the usual pions are Dirac fields, then a supersymmetric WZNW term may
exist wherein some spin-0 degrees of freedom are assigned to
supersymmetric representations {\it {other}} than the chiral multiplet!  
The nonminimal scalar superfield, a variant representation that is dual 
to the chiral superfield, has been proposed. We call such models that 
use both chiral and nonminimal superfields, chiral-nonminimal (CNM) 
models.  In the second work of reference \cite{C2}, initial steps began 
toward the description of a 4D, $N$ = 1 supersymmetric extension of 
the {\it {gauged}} WZNW term. 

In working toward this last goal we have made a search of the
literature\cite{C22}.  Although there have appeared many prior 
discussions of 4D, $N$ = 1 supersymmetric gauge theories and non-Abelian
anomalies (see also \cite{C1}), to the best of our knowledge, there has
been no discussion of the 4D, $N$ = 1 superfield action for the BGJ 
anomalies that pays {\it {special}} attention to the simplest expression 
for these anomalies.  Many of  the results given so far have explicit 
dependence on either the prepotential $V$  or $\d_G V$.  Since the 
gauge variation $\d_G V$ is a transcendental expression, the clarity 
of such formulations is lessened.  As the anomaly is always defined 
up to cohomologically trivial terms, we were led to the belief that 
the prepotential should {\it {only}}  enter via its natural superspace 
appearance as $e^V$ along with superconnections and the fermionic 
field strength.

It is thus a purpose of this work to offer an ``improved'' description
of the 4D, $N$ = 1 supersymmetric BGJ anomaly.  We begin with the
integrated form of the left-right anomalies that appear in the
(would-be) current conservation laws for the non-supersymmetric theory 
and reformulate the Wess-Zumino consistency condition in terms of two
operators, $\D$ and $\d_R$. We  also observe that the BGJ anomaly term,
defined in terms of geometrical ``monomials of the anomaly'' possesses
interesting properties with respect to these operators.  The monomials
as well as the $\D$ and $\d_R$ operators all have 4D, $N$ = 1 extensions
which we use to reconceptualize the supersymmetric problem. In the
course of our analysis, we also show how to use the results of reference
\cite{MAO} combined with a special choice of homotopy to reach our
goals.  ${~~}$ \newline

\noindent

{\bf {(II.) Algebraic Realization of the WZ Consistency Condition on the
\newline ${~~~~~~~}$Left-Right BGJ Anomaly Term}}

Our discussion starts by considering the integrated form of the 
left-right BGJ anomalies. For this purpose, we use the results of
references \cite{C, Z} with a set of gauge fields, $A_{\un a}^{(L)}$
$(  {\un a} \equiv \a \Dot \a )$, minimally coupled to purely left 
handed spinors, ${\Bar \zeta}^{\ad}$ and a set of gauge fields, 
$A_{\un a}^{(R)}$, minimally coupled to purely right handed spinors, 
${\psi}^{\a}$
$$ \eqalign{ {~~}
{\cal S}(J_L \, + \, J_R)\, &=~ {\cal S}(J_L) ~+~ {\cal S}(J_R) ~~~~, 
\cr
{\cal S}(J_L)\, &=~ \int d^4 x ~ \Big[ \,  -~ i \, {\zeta}^{\a} 
\nabla_{\un a}^{(L)} {\Bar \zeta}^{\ad} \,\,\, \Big]  \,\,~~, ~~ 
\nabla_{\un a}^{(L)} ~\equiv ~ \pa_{\un a} ~-~ i  A_{\un a}^{(L)
\, {\rm I}} t_{\rm I} ~~~~,\cr
{\cal S}(J_R)\, &=~ \int d^4 x ~ \Big[ \, -~ i \, {\Bar \psi}^{\ad} 
\nabla_{\un a}^{(R)} {\psi}^{\a} \, \Big]  ~~~, ~~ \nabla_{\un 
a}^{(R)} ~\equiv ~ \pa_{\un a} ~-~ i  A_{\un a}^{(R)
\, {\rm I}} t_{\rm I}
~~~~. }    \eqno(2.1) $$

Here $t_{\rm I}$ denotes a hermitian matrix representation of the group
generators.  We note that in supersymmetric theories, the left-right 
split is most natural since the superfields that contain the fermions
are already formulated in terms of chiral spinors. In the subsequent 
discussion we will make use of the following notational devices,
$$ \eqalign{{~~~~~~~~}
\d_G (\l) A_{\un a} &=~ \pa_{\un a} \l ~+~i \, [~ \l \, , \, A_{\un a} 
~] ~\equiv~ \pa_{\un a} \l ~+~ i \,  L_{\l} A_{\un a} ~~~, ~~~
\l ~=~ \l^I t_I~~~, \cr
\d_G (\l)  F_{\un a \un b} &=~ i\, L_{\l} F_{\un a \un b} ~~~,~~~
F_{\un a \un b} ~\equiv~ \pa_{\un a}  A_{\un b} ~-~ \pa_{\un b}  
A_{\un a} ~-~ i \, [\, A_{\un a} \, , \,  A_{\un b} ~] ~~~. 
}\eqno(2.2) $$
We emphasize that throughout the following discussion the $\l$'s are
functions of $x$.

Due to the gauge transformation properties above, it becomes possible to
define new operators that we denote by $\D$ and $\d_R$ via the equation
$$
\d_G(\l) ~=~ \D(\l) ~+~ \d_R (\l) ~~~,
\eqno(2.3)$$
where we choose 
$$ 
\d_R (\l) A_{\un a} ~=~  i \,  L_{\l} A_{\un a} ~~~, ~~~
\d_R (\l)  F_{\un a \un b} ~=~
 i \, L_{\l} F_{\un a \un b} ~~~. 
\eqno(2.4) $$
These obviously imply that
$$
\D (\l) A_{\un a} ~=~ \pa_{\un a} \l  ~~~, ~~~
\D (\l)  F_{\un a \un b} ~=~ 0 ~~~,
\eqno(2.5) $$
and show that under the action of $\D$, the connection and field
strength transform as in an abelian gauge theory. We also note that 
the operator $\D$ satisfies $\D^2 = 0$, i.e. nilpotency.  Although 
we will not exploit this  property in the present work, we believe 
this is significant since  the $\D$ operator satisfies a Poincar\' e 
lemma, suggesting an immediate relation to exact short sequences 
and topology.  As we shall see below, the decomposition in (2.3) 
is also significant for the BGJ anomaly since  actually the
$\D$-operator, not the $\d_R$-operator, determines it. We find it 
very satisfying that the nilpotent operator, reminiscent of the 
exterior derivative, plays the more fundamental role. 

We may write the left BGJ non-Abelian gauge anomaly as 
$$
{\cal S}_{\rm {BGJ}}^{(L)}(\l)  ~=~  (\frac1{48 \p^2 } ) \int d^4 
x ~ \l^{\rm I}\, {\rm G}^{\rm I}_{(L)} (A^{(L)}) ~\equiv~ 
(\frac1{48 \p^2 } ) \int d^4 x ~ \o_4^1 (\l , \, A^{(L)}, \, 
F^{(L)} \,) ~~~~~.   
\eqno(2.6)
$$ 

We have written this last function with its arguments to emphasize that 
{\it {only}} connections and field strengths are allowed to enter in the 
above construction. The function ${\rm G}^{\rm I}_{(L)} (A^{(L)})$ can
be explicitly written in the form,
$$ \eqalign{ {~} 
\l^{\rm I} {\rm G}^{\rm I}_{(L)} (A^{(L)}) &=~  {\rm {Tr}} \Big\{ 
\l \, [ \, F_{\un a \, \un b}^{(L)} {\Tilde F}{}^{\un a \, \un b ~
(L)} ~+~i \, {\Tilde F}{}^{\un a \, \un b ~(L)} \,  A_{\un a}^{(L)} 
\, A_{\un b}^{(L)} ~+~ i \, A_{\un a}^{(L)} \, {\Tilde F}{}^{\un a 
\, \un b ~ (L)} \, A_{\un b}^{(L)} \cr 
&{~~~~~~~~~~~}~+~ i \, A_{\un a}^{(L)} \, A_{\un b}^{(L)} \, {\Tilde 
F}{}^{\un a \, \un b ~(L)}~-~  \e^{\un a \, \un b \, \un c \, 
\un d} ~ A_{\un a}^{(L)}  \, A_{\un b}^{(L)} \, A_{\un c}^{(L)} \, 
A_{\un d}^{(L)} \,] \Big\} 
~~~. } \eqno(2.7) $$

In writing this, we have expressed the answer in terms of the field 
strength $F_{\un a \, \un b}$ and the dual field strength ${\Tilde 
F}{}_{\un a \, \un b } = \fracm 12 \e_{\un a \, \un b \, \un c \, \un d} 
{F}{}^{\un c \, \un d}$. The right integrated BGJ non-Abelian anomaly 
can be obtained from the left one by the replacements $ {\rm G}_{(L)} 
~\to~ - \, {\rm G}_{(R)}$, $A_{\un a}^{(L)} ~\to ~  A_{\un a}^{(R)}$
and $F_{\un b \,\un c}^{(L)}   ~\to ~ F_{\un b \, \un c}^{(R)}$. The 
leading term in (2.6, 2.7) has exactly the same form as that for the 
Abelian anomaly with the difference that an extra factor of the group
generator $t^{\rm  I}$ (contained in $\l$) is present under the trace
operation. 

The BGJ non-Abelian gauge anomaly term (2.6, 2.7) is by definition
a solution of the Wess-Zumino consistency condition,
$$
\D (\l_1) \, {\cal S}_{\rm {BGJ}}^{(L)}(\l_2 )  ~-~ 
\D (\l_2) \, {\cal S}_{\rm {BGJ}}^{(L)}(\l_1 ) ~=~  i \,
{\cal S}_{\rm {BGJ}}^{(L)}([\, \l_1 \, , \, \l_2 \, ] )
~~~,
\eqno(2.8) $$ 
where we have used the $\D(\l)$ operator in place of the more
traditional $\delta_G(\l)$ operator. We note that using the 
$\delta_G$ operator leads to the same expression, but with a 
minus sign on the RHS of (2.8). It is also of interest to note 
that
$$
\d_R (\l_1) \, {\cal S}_{\rm {BGJ}}^{(L)}(\l_2 )  ~-~ 
\d_R (\l_2) \, {\cal S}_{\rm {BGJ}}^{(L)}(\l_1 ) ~=~ - i 2\,
{\cal S}_{\rm {BGJ}}^{(L)}([\, \l_1 \, , \, \l_2 \, ] )
~~~,
\eqno(2.9) $$ 
which differs from the WZ consistency condition in (2.8) by a factor of
$(-2)$ on the right hand side. Analogous identities hold for the right 
BGJ anomaly $S_{\rm BGJ}^{(R)}$.

The function $\o_4^1 $ in (2.6) can  be organized according to the
powers of the connections that enter its different terms. So that 
we have
$$
\o_4^1 (\l , \, A^{(L)}, \, F^{(L)} \,) ~\equiv~ {\cal A}_0 (\l)
~+~ i \, {\cal A}_2 (\l) ~-~  {\cal A}_4 (\l) ~=~ \sum_{\ell = 
0}^{2}(i)^{\ell} {\cal A}_{2\ell}(\l) 
~~~,
\eqno(2.10) $$
where the subscripts of ${\cal A}_{2l}$ denotes the powers of the
connection that enter, i.e.
$$ \eqalign{ 
{\cal A}_0 (\l) &=~ {\rm {Tr}}  \Big\{ \l \,  F_{\un a \, \un b} {\Tilde 
F}{}^{\un a \, \un b} \, \Big\}  ~~~, \cr
{\cal A}_2 (\l) &=~ {\rm {Tr}}  \Big\{\l \, [ A_{\un a} \, A_{\un b} \, 
{\Tilde F}{}^{\un a \, \un b} \,+\, A_{\un a} \, {\Tilde F}{}^{\un a 
\, \un b} \, A_{\un b} \,+\, {\Tilde F}{}^{\un a \, \un b}\, A_{\un a} 
\, A_{\un b} \, ] \, \Big\} ~~~, \cr  
{\cal A}_4 (\l) &=~ \e^{\un a \, \un b \, \un c \, \un d} ~ {\rm {Tr}}  
\Big\{\l \,A_{\un a} A_{\un b} A_{\un c} A_{\un d} \, \Big\} ~~~,
} \eqno(2.11)$$
and we dropped the $L$ superscript for notational convenience.  For
reasons that will become clear later, we call the ${\cal A}_i$'s, 
``the basis monomials of the anomaly.''

It is instructive (especially in view of our goal to treat the 4D, $N$ 
= 1 supersymmetric case) to ask precisely how the WZ consistency condition
gets satisfied using the monomials. In order to do this, it is useful to
introduce a further notational device.  Let the symbol $\Big\{\Big\}^{\un 
a \, \un b \, \un c \, \un d}$ be defined by
$$ 
\Big\{  \Big\}^{\un a \, \un b \, \un c \, \un d} ~\equiv~ \e^{\un 
a \, \un b \, \un c \, \un d} ~ {\Tilde {\rm {Tr}}}  \Big\{ ~ \Big\} 
~~~, 
\eqno(2.12) $$ 
where, given any matrices ${\cal X}$, ${\cal Y}$ and ${\cal Z}$, we
define
$$
{\Tilde {\rm {Tr}}}  \Big\{ ~ {\cal X} \, \l_0 \, {\cal Y}
\, \l_1 \, {\cal Z}~ \Big\} ~\equiv ~ {\rm {Tr}}  \Big\{ ~ 
{\cal X} \, \l_0 \, {\cal Y} \, \l_1 \, {\cal Z}~ \Big\} ~-~
{\rm {Tr}} \Big\{ ~ {\cal X} \, \l_1 \, {\cal Y}
\, \l_0 \, {\cal Z}~ \Big\}
~~~. \eqno(2.13)
$$

A simple set of calculations reveals
$$ \eqalign{ {\,}
\D (\l_1) \, {\cal A}_0 (\l_2 ) ~-~ \D (\l_2) \, {\cal A}_0 (\l_1 ) &=~
0 ~~~, {~~~~~~~~~} \cr
\D (\l_1) {\cal A}_2 (\l_2) ~-~ \D (\l_2) {\cal A}_2 (\l_1) &=~ {\cal
A}_0 ([ \,\l_1  \, , \, \l_2 \,]) \,+\, i \, \Big\{ \, \l_1 \, \l_2 \, 
A_{\un a}   F_{\un b \un c} A_{\un d} \,+\, \frac 12 [ ~ A_{\un a} \l_1 
A_{\un b} \l_2 \, \cr  
&{~~~~}+ \, \l_1 A_{\un a} \l_2  A_{\un b} \, - \, \l_1 A_{\un a}
A_{\un b}\l_2 ~] F_{\un c \un d} ~ {\Big\} }^{\un a \, \un b \, \un c \, 
\un d} ~~~, \cr
\D (\l_1) {\cal A}_4 (\l_2) ~-~ \D (\l_2) {\cal A}_4 (\l_1) &=~  i \,
{\cal A}_4 ([ \,\l_1 \, , \, \l_2 \,]) ~+~  {\cal A}_2 ([ \,\l_1 \, 
, \, \l_2 \,]) {~~~~~~~~~} \cr
&{~~~~}-~ \Big\{ \, \l_1 \, \l_2 \, A_{\un a}   F_{\un b \un c} A_{\un 
d} \,+\, \frac 12 [ ~ A_{\un a} \l_1 A_{\un b} \l_2 \, \cr  
&{~~~~}+ \, \l_1 A_{\un a} \l_2  A_{\un b} \, - \, \l_1 A_{\un a}
A_{\un b}\l_2 ~] F_{\un c \un d} ~ {\Big\} }^{\un a \, \un b \, \un c \, 
\un d}   ~~~.} \eqno(2.14) $$ 
In writing these results, we have neglected total divergences. It is
also useful to note that we have repeatedly used the identities
$$ 
\pa_{\un a} A_{\un b} ~=~ \frac 12 F_{\un a \, \un b} ~+~ i \, A_{\un 
a} A_{\un b} ~~~, ~~~ \pa_{\un a} F_{\un b \, \un c}  ~=~ - i \,
F_{\un a \, \un b} A_{\un c}  ~+~ i \,  A_{\un a} F_{\un b \, \un c} 
~~~. \eqno(2.15) $$
which are valid under the $\Big\{  \Big\}^{\un a \, \un b \, \un c \, 
\un d}$ symbol.  It is seen that upon introducing constants $a_0$, $a_2$ 
and $a_4$, we obtain
$$ \eqalign{ {~~} 
&\D (\l_1) \Big[ ~ \sum_{ \ell = 0}^{2} a_{2 \ell} {\cal A}_{2 \ell}
(\l_2) ~\Big]  ~-~ \D (\l_2) \Big[ ~ \sum_{ \ell = 0}^{2} a_{2 \ell} 
{\cal A}_{2 \ell} (\l_1) ~\Big]  ~=~ \cr
&{~~~~} a_2 {\cal A}_0 ([ \,\l_1 \, , \, \l_2 \,]) ~+~ a_4  {\cal A}_2 
([ \,\l_1 \, , \, \l_2 \,])  ~+~ i \, a_4 \, {\cal A}_4 ([ \,\l_1 \, , 
\, \l_2 \,])  \cr
&{~~~~}-~ (a_4 \, - \, i \,  a_2 ) \, \Big\{ ~ \Big[ ~ - 2\, A_{\un a} 
 \,\l_1 \,  \l_2 \, A_{\un b}  ~+~ \l_1 A_{\un a} \l_2  A_{\un b}
~-~ \l_1 A_{\un a} A_{\un b}\l_2 ~ \Big]  F_{\un c \un d} \Big\}^{\un 
a \, \un b \, \un c \, \un d} ~~~. }\eqno(2.16) $$
Imposing the WZ consistency result (2.8) upon this last equation, 
we find that the constants must satisfy $a_0 = - i a_2 = - a_4$ 
which up to an overall normalization reproduces (2.6,2.7).  It is 
also of note that each of the monomials {\it {separately}} and 
{\it {exactly}} (i.e. no total divergences are dropped) satisfies 
the equation
$$
\d_R (\l_1) \, {\cal A}_{2 \ell} (\l_2 )  ~-~ \d_R (\l_2) \, {\cal 
A}_{2 \ell} (\l_1 ) ~=~ - i 2\, {\cal A}_{2 \ell}([\, \l_1 \, , \, 
\l_2 \, ] ) ~~~.
\eqno(2.17) 
$$

This approach also emphasizes that it is the abelian part of the gauge 
field transformation law that determines the form of the BGJ anomaly 
when working in the basis defined by the anomaly monomials in (2.11).  
The condition in (2.17) does not lead to algebraic relations among the 
$a_{2 \ell}$ coefficients.

It is also interesting to investigate how the WZ consistency condition
is satisfied when the anomaly is given as 
$$ 
\eqalign{ {~~~~~~~~~~} 
\l^I {\rm G}_{(L)}^I (A^{(L)}) &=~  \e^{\un a \, \un b 
\, \un c \, \un d} \, {\rm {Tr}}  \Big\{  \l \, [  \pa_{\un a} \,(\,
A_{\un b} \, F_{\un c \, \un d} \,+\,  i \, A_{\un b} \,
A_{\un c} \,  A_{\un d}  ~) ~] ~ \Big\} \cr 
&\equiv ~ {\cal B}_1 (\l) ~+~  i\, {\cal B}_3 
(\l) ~~~. } 
\eqno(2.18) $$
Here ${\cal B}_1$ denotes the term linear in the gauge field and ${\cal  
B}_3$ denotes the term cubic in the gauge field.  Once again the leading 
term in (2.18) is of the form of the Abelian anomaly except for the 
presence of $\l$ under the trace operation. In this way of writing the
anomaly we find that the WZ consistency condition leads to
$$
\d_R (\l_1) \, {\cal S}_{\rm BGJ}^{(L)} (\l_2 )  ~-~ 
\d_R (\l_2) \, {\cal S}_{\rm BGJ}^{(L)} (\l_1 ) ~=~ - i \,
{\cal S}_{\rm BGJ}^{(L)} ([\, \l_1 \, , \, \l_2 \, ] )
~~~,
\eqno(2.19) $$
as a trivial consequence of the rigid transformations (2.4), and
$$
\D (\l_1) \, {\cal S}_{\rm {BGJ}}^{(L)}(\l_2 )  ~-~ 
\D (\l_2) \, {\cal S}_{\rm {BGJ}}^{(L)}(\l_1 ) ~=~  0
~~~,
\eqno(2.20) $$ 
as the only non--trivial condition.

The condition (2.20) can be used to determine algebraically the
non-abelian anomaly in the form (2.18). In fact, we can start with
the definition
$$
\o_4^1 ~=~ b_1 \, {\cal B}_1 ~+ b_3 \, {\cal B}_3
\eqno(2.21)
$$
where $b_1$ and $b_3$ are generic constants. To show that solutions 
to equation (2.20) exist in this form we note that up to total 
derivatives
$$ \eqalign{ {\,~~~~}
\D (\l_1) {\cal B}_1 (\l_2) ~-~ \D (\l_2) {\cal B}_1 
(\l_1) &=~  \e^{\un a \, \un b \, \un c \, \un d} \, {\Tilde 
{\rm {Tr}}} \Big\{ (  \pa_{\un a} \l_1 ) \, (\, \pa_{\un b} 
\l_2 ) \, F_{\un c \, \un d} ~ \Big\} {~~~,~~~~~~}  \cr
\D (\l_1) {\cal B}_3 (\l_2) ~-~ \D (\l_2) {\cal B}_3 (\l_1) 
&=~ i \, \e^{\un a \, \un b \, \un c \, \un d} \, {\Tilde 
{\rm {Tr}}} \Big\{  (  \pa_{\un a} \l_1 ) \, (\, \pa_{\un 
b} \l_2 ) \, F_{\un c \, \un d} ~) ~ \Big\} {~~~,~~~~~~} 
}\eqno(2.22) $$
where we have used the Bianchi identities (2.15).  As long as the
condition $b_3 = ib_1$ is valid, we see that (2.20) is satisfied.

We note that the non-Abelian consistency condition as
encoded in the  $\d_R$-equation which does not lead to any algebraic
condition on the $b_i$ coefficients, whereas the $\D$-equation simply
yields an Abelian-like condition which fixes the constants $b_1$ 
and $b_3$ up to an overall normalization factor. Again, the structure 
of the anomaly is completely  determined by the nilpotent 
$\D$-operator.

Since it is our goal to study the possibility of a supersymmetric
generalization of (2.7), it behooves us to make one final set of
notational changes to facilitate the use of {\it {Superspace}}
\cite{SS} conventions.  We note
$$
A_{\un a} ~=~ A_{\a \Dot \a} ~~~, ~~~
\e^{\un a \, \un b \, \un c \, \un d} ~=~  i \,  
[ ~ C^{\a \d} C^{\b \g} C^{ \Dot \a \Dot \b } C^{\Dot \g  \Dot \d}
- C^{\a \b} C^{\g \d} C^{ \Dot \a \Dot \d } C^{ \Dot \b \Dot \g } ~]
~~~, 
\eqno(2.23)
$$
$$
F_{\un a \, \un b} ~=~ [\,  C_{\Dot \a \Dot \b} f_{\a \b} ~+~
C_{\a \b} {\bar f}{}_{\Dot \a \Dot \b}  \,] ~~~,~~~
{\Tilde F}{}_{\un a \, \un b} ~=~  i \, [\,  C_{\Dot \a \Dot 
\b} f_{\a \b} ~-~ C_{\a \b} {\bar f}{}_{\Dot \a \Dot \b} \,]
 ~~~,
\eqno(2.24)
$$
so that the final form in which we write (2.6, 2.7) is
$$ 
\eqalign{ 
{~}{\cal S}_{\rm {BGJ}}^{(L)}(\l) ~=~  (\frac1{24 \p^2 } ) \int d^4 x  
~ {\rm {Tr}}  \Big\{ \l \, [ \, &i \, f^{\a \, \b} f_{\a \, \b} ~-~ i  
{\bar f}{}^{\Dot \a \, \Dot \b}  {\bar f}{}_{\Dot \a \, \Dot \b} ~+~ 
\frac 12 f^{\a \, \b} \, A_{\a}{}^{\Dot \b} \, A_{\b \Dot \b} 
{~~~~~~~~}\cr
&+~ \frac 12 \, A_{\a}{}^{\Dot \b} \, f^{\a \, \b} \, A_{\b \Dot \b} 
~+~ \frac 12 \, A_{\a}{}^{\Dot \b} \, A_{\b \Dot \b} \, f^{\a \, \b} \cr 
&-~ \frac 12 {\bar f}{}^{\Dot \a \, \Dot \b} \, A^{\a}{}_{\Dot \a} \, 
A_{\a \Dot \b} ~-~ \frac 12  \,  A^{\a}{}_{\Dot \a} \, {\bar f}{}^{\Dot 
\a \, \Dot \b}   \, A_{\a \Dot \b} \cr
&-~ \frac 12  \,  A^{\a}{}_{\Dot \a} \, A_{\a \Dot \b} \, {\bar
f}{}^{\Dot \a \, \Dot \b} ~-~ \frac{i}{2} A^{\a \Dot \a} \, A_{\a \Dot 
\b} \, A^{\b \Dot \b} \, A_{\b \Dot \a} \cr
&+~ \frac{i}{2} A^{\a \Dot \a} \, A_{\b \Dot \a} \, A^{\b \Dot \b} \, 
A_{\a \Dot \b}  \,] ~ \Big\} ~~~, } 
\eqno(2.25) $$
whereas equation (2.18) takes the form
$$ 
\eqalign{ 
{~}{\cal S}_{\rm {BGJ}}^{(L)}(\l) ~=~ (\frac1{24 \p^2 } ) \int d^4 x ~
{\rm {Tr}}  \Big\{ \l \, \pa^{\un a} \Big( \, &i \, A_{\a}{}^{\Dot \b}
{\bar f}{}_{\Dot \a \, \Dot \b} ~-~ i A^{\b}{}_{\Dot \a}{f}{}_{\a \, \b} 
~+~ \frac 12  A_{\a}{}^{\Dot \b} \, A^{\b}{}_{\Dot \b} \, A_{\b \Dot \a} 
\cr
&{~~~~~} ~-~  \frac 12  A^{\b}{}_{\Dot \a} \, A_{\b}{}^{\Dot \b} \,
A_{\a \Dot \b} \, \Big) ~ \Big\}
~~~. } \eqno(2.26) $$
A superfield action for the anomaly should contain the terms in either
(2.25) or (2.26) at a minimum.
${~~~}$\newline

\newpage

\noindent

{\bf {(III.) Preliminaries for 4D, $N$ = 1 Supersymmetric BGJ Anomaly
Term}}

Having completed the discussion of relevant structures in the
non-supersymmetric case, the next obvious step is to consider analogous
structures in the supersymmetric extensions.  The superspace Yang-Mills 
covariant superderivative $\nabla_{\un A} = D_{\un A} \,-\, i \G_{\un A}$ 
(where $\G_{\un A}$ is a matrix in the Lie algebra of the gauge group and 
$\nabla_{\un A} \equiv (\nabla_{\a}, \, {\nabla}{}_{\Dot \a}, \,
\nabla_{\un  a}$)) is totally expressed in the chiral representation in
terms of a pseudoscalar hermitian-matrix general superfield $V$ as
$$
\nabla_{\a} ~\equiv~ e^{-V} \, D_{\a} e^V ~~~~,~~~~ \nabla_{\Dot \a} 
~\equiv~  {\Bar D}_{\Dot \a} ~~~~,~~~~ \nabla_{\un a} ~\equiv~ - 
i \, \Big\{ ~ \nabla_{\a} \, , \, \nabla_{\Dot \a} ~\Big\}
~~~,
\eqno(3.1) $$
and the spinorial field strengths are given by ($ {\Bar \G}{}_{\Dot \a}
= -( \G_{\a})^*$ )
$$ \eqalign{ {~~}
W_{\a} &=~ {\Bar D}{}^2 \G_{\a} ~=~ i {\Bar D}{}^2 ( e^{-V} \, D_{\a}
e^V \,) ~~~, ~~~ {\Bar W}{}_{\Dot \a} ~=~ {D}{}^2 {\Bar \G}{}_{\Dot \a} 
~=~ i {D}{}^2 (e^{V} \, {\Bar D}_{\Dot \a} e^{-V} \,)  ~~~.
}\eqno(3.2)
$$
Here we use {\it {Superspace}} \cite{SS} notations and conventions
supplemented by superspace conjugation rules on the derivatives, 
$(D_{\a})^* \,=\, - {\Bar D}{}_{\Dot \a}$, $({D}{}^{\a})^* \,=\, 
{\Bar D}{}^{\Dot \a}$ and $(\pa_{\un a})^* \, = \, \pa_{\un a}$.

From the results in (3.1) and (3.2) we see that
$$
{\Bar D}{}_{\Dot \a} \G_{\a} ~=~ i \, \G_{\un a} ~~~~,~~~~
{\Bar D}{}^{\Dot \a} \G_{\un a} ~=~ - i 2 \, W_{\a} ~~~~, 
\eqno(3.3) 
$$
$$
{\Bar D}{}_{\Dot \a} W_{\a} ~=~ 0 ~~~~, ~~~~ \nabla^{\a} W_{\a} ~+~
{\Bar D}{}^{\Dot \a} \, ( e^{-V}\, {\Bar W}{}_{\Dot \a}  \, e^V ) ~=~ 0
~~~~.
\eqno(3.4) $$
where the covariant derivative acting on  superfields in the adjoint 
representation of the gauge group is defined as $\nabla^{\a} W_{\a}
\equiv D^{\a} \, W_{\a} - i \, \{ \, \G^{\a} \, , \, W_{\a} \, \}$. 
Most of the following equations are familiar from the literature,
$$
\eqalign{ {~~~~}
 D_{\a} \, e^V &=~ - i \, e^V \, \G_{\a} ~~~,~~~  D_{\a} \, e^{-V} ~=~ 
i \, \G_{\a}\, e^{-V}  ~~~,  \cr 
{\Bar D}{}_{\Dot \a} \, e^V &=~  i \, {\Bar \G}{}_{\Dot \a} \, e^V
~~~,~~~ {\Bar D}{}_{\Dot \a} \, e^{-V} ~=~ - i \, e^{-V} \, {\Bar 
\G}{}_{\Dot \a} ~~~,  \cr
\pa_{\un a}\,  e^V &=~ - i \, e^V \, \G_{\un a} ~+~ i \, {\Bar
\G}{}_{\un a} \, e^V ~~~,~~~ \pa_{\un a} \, e^{-V} ~=~ - i \, e^{-V} \,
{\Bar\G}{}_{\un a} ~+~ i \, \G{}_{\un a} \, e^{-V} ~~~, } \eqno(3.5)$$ 
but to our knowledge the last two have not appeared before.

The results of (3.3) and (3.5) are the supersymmetric analogs of the
first result given in (2.15) for the non-supersymmetric theory.  In 
both cases, the equations tell us how to calculate the first derivatives 
of gauge-variant objects in terms of other geometrical quantities, the
superconnections. Similarly, the results in (3.4) are known to be the
supersymmetric analogs of the second result given in (2.15) for the
non-supersymmetric theory.  In both cases these are the Bianchi
identities. 

The infinitesimal gauge transformations of $e^V$, ${\G}_{\a}$ and
$W_{\a}$ are given by
$$ \eqalign{ {~~~}
\d_G(\L) \, e^V &= ~ i \, [ ~ {\Bar \L} \, e^V ~-~ e^V \, \L ~ ] ~~~,
~~~ \d_G(\L) \, e^{-V} ~= ~ i \, [ ~ \L \, e^{-V} ~-~ e^{-V} \, {\Bar
\L} ~ ] ~~~, \cr 
\d_G(\L) \, \G_{\un A} &=~ D_{\un A} \, \L ~+~ i \, L_{\L} \, \G_{\un 
A} ~~\,~,~~~~ \d_G(\L) \, W_{\a} ~=~ i \, L_{\L} \, W_{\a} ~~~,
 } \eqno(3.6) $$
where the gauge parameter superfields satisfy 
${\Bar D}{}_{\Dot \a} \L \, = \, D_{\a} \, {\Bar \L} \, = \, 0$.

In the bosonic case we defined the operator $\D$ and saw that the WZ 
consistency condition can be entirely reformulated in terms of this 
operator. The $\D$-operator has a natural extension to a supersymmetric 
YM theory.  In this case it can be split into a sum of a holomorphic 
operator $\D_1$ and its antiholomorphic conjugate ${\Bar \D}{}_1$ 
according to 
$$
\D ~=~ \D_1 ~+~ {\Bar \D}{}_1  ~~~,
\eqno(3.7) $$
where both $\D_1$ and ${\Bar \D}{}_1$ annihilate all field strengths and
factors $e^V$ and $e^{-V}$.  However, they act on superconnections as
$$
\D_1 \, \G_{\a} ~=~ D_{\a} \, \L ~~~,~~~ {\Bar \D}{}_1 \, \G_{\a} ~=~ 
0 ~~~,~~~ \D_1 \, \G_{\un a} ~=~ \pa_{\un a} \, \L ~~~,~~~ {\Bar
\D}{}_1 \, \G_{\un a}  ~=~ 0 ~~~.
\eqno(3.8) $$
In common with its non-supersymmetric precursor, the holomorphic
supersymmetric operator $\D_1$ and its antiholomorphic conjugate 
${\Bar \D}{}_1$ satisfy, $(\D_1)^2 = ({\Bar \D}{}_1)^2 = \D{}_1 
{\Bar \D}{}_1 = {\Bar \D}{}_1 \D{}_1 = 0$.

In a similar manner, the supersymmetric operator $\d_R$ can also be
split into the sum of a holomorphic operator $\d^1_R$ and anti-holomorphic 
operator ${\bar \d}^1_R$,
$$
\d_R ~=~ \d^1_R ~+~ {\bar \d}{}^1_R ~
~~,
\eqno(3.9) $$
where 
$$ \eqalign{
\d^1_R \, e^V &=~ - i \, e^V \, \L ~~~~,~~~~ \d^1_R \, e^{-V} ~=~ 
i \, \L \, e^{-V} ~~~~, \cr
\d^1_R \, \G_{\un A} &=~  i \, L_{\L} \, \G_{\un A} ~~~\,~,~~~\,~ \d^1_R
\, W_{\a} ~=~ i \, L_{\L} \, W_{\a} ~~~, \cr
\d^1_R \, {\Bar \G}{}_{\un A} &=~ 0 ~~~~~~~~~~\,~~,~~~~ \d^1_R \,
{\Bar W}{}_{\Dot \a} ~=~ 0 ~~~. \cr
} \eqno(3.10) $$
We also note that ``tilde-variables'' defined by
$$
{\Tilde \G}{}_{\Dot \a} ~\equiv ~ e^{-V} \, {\Bar \G}{}_{\Dot \a} \, 
e^V ~~~,~~~ {\Tilde \G}{}_{\un a} ~\equiv ~ e^{-V} \, {\Bar \G}{}_{\un 
a} \, e^V ~~~,~~~ {\Tilde W}{}_{\Dot \a} ~\equiv ~ e^{-V} \, {\Bar 
W}{}_{\Dot \a} \, e^V ~~~, 
\eqno(3.11) $$
only transform under the action of $\d^1_R$ according to eqs. (3.10). 
We will call ``holomorphic'' any quantity which manifests this behavior 
under the $\d^1_R$ transformation. ${~~~}$\newline

The superfield form of ${\cal S}(J_L + J_R)$ is usually assumed to be of 
the form
$$
{\cal S}_{C^2} (J_L + J_R) ~=~ \int d^8 Z  ~ [ ~ {\Bar \Phi}{}_+
e^{V^{(R)}}  {\Phi}{}_+ ~+~ \Phi_- e^{- V^{(L)} } {\Bar \Phi}{}_- 
~] ~~~,
\eqno(3.12)
$$
so that the spinor $\zeta_{\a}$ are contained in $\Phi_-$ and the
spinors $\psi_{\a}$ are contained $\Phi_+$. We use the notation $d^8 Z
\equiv d^4 x d^2 \q d^2 {\bar \q}$.   For a finite gauge transformation,
the matter superfields $\Phi_-$ and $\Phi_+$ and the Yang-Mills gauge 
superfields $V^{(L)}$ and $V^{(R)}$ transform as
$$
\eqalign{ {~~~~~~~} ( {{\Phi}}{}_+ )'  & = ~ \exp[ i {\L}{}^{(R) 
\, I} t_I ] \, {{\Phi}}{}_+  ~~~~\,~,~~~~\,~ ({\Bar \Phi}{}_+ )' 
~ = ~  {\Bar \Phi}_+\, \exp[-i {\Bar{\L}}{}^{(R) \, I} t_I ]  
~~~,  \cr
( {\Bar {\Phi}}{}_- )' & = ~ \exp[ i {\Bar \L}{}^{(L) \, I} t_I ]
\, {\Bar {\Phi}}{}_- ~~~~~~,~~~~\,~ (  {\Phi}{}_- )' ~= ~
{\Phi}{}_-\, \exp[- i \L{}^{(L) \, I} t_I ] ~~~, \cr
(e^{- V^{(L)} })' &=~ e^{i \L^{(L)} } \, e^{- V^{(L)} } \,  e^{-i 
{\Bar \L}{}^{(L)} }~~~\,,  \,~~~ (e^{V^{(R)} })' ~=~ e^{i {\Bar \L}{
}^{(R)} } \,  e^{V^{(R)} } \,  e^{-i {\L} {}^{(R)} }~~~. 
 }\eqno(3.13) $$
We shall construct the anomaly associated with either Left or Right
gauge group.

\noindent

{\bf {(IV.) The 4D, $N$ = 1 Supersymmetric BGJ Anomaly Term in the
\newline ${~~~~~~~}$ Wess-Zumino Gauge}}

The BGJ anomaly as written in eq. (2.18) strongly suggests that any 
supersymmetric extension can be written proportional to $D_{\un A}
\, \L$.  On the basis of dimensional analysis (in mass units $dim[V] 
= 0$, $dim[\G_{\a}] = \frac 12$, $dim[ \G_{\un a}] = 1$ and $dim[W_{
\a}] = \frac 32$) one can then write all possible monomials proportional 
to $D_{\un A} \, \L$ times connections $\G_{\un A}$, field strengths
$W_{\a}$, $\bar{W}_{\Dot \a}$ or their derivatives and generic functions
of $e^V$.  The physical bosonic content of any single monomial can
be easily found by linearizing in $V$ and reducing in components in the 
Wess-Zumino gauge, defined by the following conditions 
$$ 
V | ~=~ 0 ~~~,~~~ D_{\a} V | ~=~ 0 ~~~,~~~ D^2 V | ~=~ 0 ~~~.
\eqno(4.1) 
$$
In this gauge, the reality condition $\L| = \bar{\L}| \equiv \l$ holds 
for the gauge parameters.  Moreover, due to the chirality constraint, 
they satisfy $\bar{D}_{\Dot \a} \, D_{\a} \, \L | = D_{\a} \, \bar{D
}_{\Dot \a} \bar{\L} | = i \, \pa_{\un a} \l$.  Performing the reduction 
for all the geometrical objects and keeping only the physical bosonic
components we have found that the only monomial structures proportional
to $D_{\un A}\L$ and linear in $V$, which give contributions to 
the bosonic physical sector are
$$ \eqalign{ {~}
\int \, d^8Z \, {\cal C}_1 &\equiv~ \int \, d^8Z \, {\rm {Tr}}
\Big\{ D^{\a} \, \L \, \{ W_{\a} \, , \, V \} \, \Big\} ~+~ {\rm {h.c.}}
~ \rightarrow  - \int \, d^4 x \, {\cal B}_1 ~~~, \cr
\int \, d^8Z \, {\cal C}_2 &\equiv~ \int \, d^8Z \, {\rm {Tr}} \Big\{ 
\pa_{\un a} \, \L \,  [ \G^{\a} \, , \, [ \bar{\G}^{\Dot \a} \, , \, 
V] \, ] \Big\} ~+~ {\rm {h.c.}} ~ \rightarrow  2\, i \, \int \, 
d^4 x \, {\cal B}_3  ~~~, \cr 
\int \, d^8Z \, {\cal C}_3 &\equiv~ \int \, d^8Z \, {\rm {Tr}}
\Big\{ D^{\a} \, \L \, \{ \G_{\un a} \, , \, [ \bar{\G}^{\Dot \a} \, , 
\, V ] \} \Big\} ~+~ {\rm {h.c.}} ~ \rightarrow \int \, d^4 x \, [\,
{\cal B}_1 ~+~2  \, i \, {\cal B}_3 \,]  ~~~, \cr 
\int \, d^8Z \, {\cal C}_4 &\equiv~ \int \, d^8Z \, {\rm {Tr}}
\Big\{ D^{\a} \, \L \, \{ \bar{\G}_{\un a} \, , \, [ \bar{\G}^{\Dot \a} 
\,, \, V] \} \Big\} ~+~ {\rm {h.c.}} ~ \rightarrow -4 \, i \, 
\int \, d^4 x \, {\cal B}_3  ~~~, \cr \
\int \, d^8Z \, {\cal C}_5 &\equiv~ i \, \int \, d^8Z \, {\rm {Tr}} 
\Big\{ D^{\a} \, \L \, \{ \pa_{\un a} \, \bar{\G}^{\Dot \a} \, , 
\, V\} \Big\} ~+~ {\rm {h.c.}} ~ \rightarrow \int \, d^4 x \, [\, 
{\cal B}_1 ~+~ 2 \,  i \, {\cal B}_3 \,]   ~~~,  \cr 
\int \, d^8Z \, {\cal C}_6 &\equiv~ i \, \int \, d^8Z \, {\rm {Tr}} 
\Big\{ D^{\a} \, \L \, \{ \bar{D}^{\Dot \a} \, \bar{\G}_{\un a} 
\, , \,  V \}  \Big\} ~+~ {\rm {h.c.}} ~ \rightarrow \int \, d^4 x \, 
[\, {\cal B}_1 ~+~ 2 \,  i \, {\cal B}_3 \,]  ~~~, } $$
$$ \eqalign{ {~}
\int \, d^8Z \, {\cal C}_7 &\equiv~ \int \, d^8Z \, {\rm {Tr}}
\Big\{ D^{\a} \, \L \, \Big( \{ \{  \G_{\un a} \, , \, \bar{\G}^{\Dot 
\a} \} \, , \, V \}  ~-~ \{ \{ \bar{\G}_{\un a} \, , \, \bar{\G}^{\Dot 
\a} \} \, , \, V \} \Big) \Big\} ~+~ {\rm {h.c.}}  \cr
& ~~~~~~~~~~~~~~~~~~~~~~~~~~ 
\rightarrow  \int \, d^4 x \, [\,-3{\cal B}_1 ~+~ 2 \,
i \, {\cal B}_3 \,]  ~~~. }\eqno (4.2) $$
It is easily seen that one can find many suitable linear combinations 
of the superfield monomials (the ${\cal C}$'s) whose component reduction
gives the correct bosonic combination $(- {\cal B}_1 - i {\cal B}_3)$.
For example we conclude that, within the WZ gauge,
$$ 
{~}{\cal S}_{\rm {BGJ}}(\l) ~=~  (\frac1{48 \p^2 } ) \,  \int \, d^8Z ~
[ ~ {\cal C}_1 \,- \, \frac 12 \, {\cal C}_2 ~+~ {\rm {h.c.}} ~ ] ~~~, 
\eqno (4.3) $$
contains the proper expression for the anomaly. Therefore, the
expression for the supersymmetric BGJ anomaly necessarily contains this 
particular linear combination up to extra terms which could be required 
in order to satisfy the supersymmetric WZ consistency condition but 
whose physical bosonic sector is cohomologically trivial.

As shown in eq. (4.2), the reduction to the WZ gauge allows only for the 
determination of the linear $V$ dependence of the supersymmetric anomaly 
on the prepotential. In the next section, by exploiting the general 
approach of Ref. \cite{MAO}, we determine the simplest structure of 
the supersymmetric monomials as functions of the geometric objects of
the theory.     ${~~~}$\newline

\noindent
{\bf {(V.) The Holomorphic 4D, $N$ = 1 SUSY BGJ Anomaly Term, the 
\newline ${~~~~~~~}$MAO Formalism and the Minimal Homotopy}}

Outside of the Wess-Zumino gauge it is reasonable to seek a
supersymmetric extension of the BGJ anomaly in the form,
$$
{\cal S}_{BGJ}(\L, \, {\Bar \L} ) ~=~  [ ~ {\Tilde S}_{BGJ} 
(\L) ~+~ {\rm {h.\, c.}}~ ]  ~~~.
\eqno(5.1) $$
We shall call the action ${\Tilde S}_{BGJ} (\L)$ the ``holomorphic 
BGJ anomaly action'' and its hermitian conjugate the ``anti-holomorphic 
BGJ anomaly action.''  This action is subject to two conditions: 
(a.) it must satisfy the superfield  WZ consistency condition and  
(b.) it must not be cohomologically trivial.  In terms of the $\D_1$ 
and $\d^1_R$ operators, the superfield WZ consistency  condition 
for the purely holomorphic BGJ anomaly action (and equivalently, 
for the purely anti-holomorphic BGJ anomaly action) is
$$
[ \, {\d}{}^1_R (\L_1) \,+\, {\D}{}_1 (\L_1) \,]  {\Tilde {\cal S}}{}
_{\rm {BGJ}}(\L_2 )  ~-~ [ \, {\d}{}^1_R (\L_2) \,+\, {\D}{}_1 
(\L_2) \,] \, {\Tilde {\cal S}}{}_{\rm {BGJ}}(\L_1 ) ~=~  ~~~~~~~~~~~~~
$$
$$
{~~~~~~~~~~~~~~~~~~~~~~~~~~~~~~~~~~~~~~~~~~~~~~~~~~~~~~~~~~~~~} - \, i
\,  {\Tilde {\cal S}}{}_{\rm {BGJ}}([\, \L_1 \,  , \, \L_2 \, ] ) ~~~,
\eqno(5.2) $$
and also a second {\it {independent}} condition given by
$$ \eqalign{ {~~}
&Re \Big\{\,[ \, {\bar \d}{}^1_R (\L_1) \,+\, {\Bar {\D}}{}_1 (\L_1) \,] 
{\Tilde {\cal S}}{}_{\rm {BGJ}}(\L_2 )  ~-~ [ \, {\bar \d}{}^1_R (\L_2)
\,+\,  {\Bar {\D}}{}_1 (\L_2) \,] \, {\Tilde {\cal S}}{}_{\rm {BGJ}}(\L_1) 
\Big\} ~=~ 0 ~~~.
}\eqno(5.3) $$

The supersymmetric BGJ anomaly is expressible in terms of a real super
4-form $F_{ABCD}$. The superspace geometry of all 4D, $N$ = 1 irreducible 
super $p$-forms was established many years ago \cite{PFM}. Components 
of the real  4-form satisfy the following constraints,
$$
F_{\a \, \b \, \g ~ \un D} ~=~ F_{\ad ~ \b \, \g \, \un D} ~=~
F_{\ad \, \b \, \un c \, \un d} ~=~ 0 ~~~,~~~ F_{\a \, \b \, \un c 
\, \un d} ~=~  C_{\Dot \g \Dot \d} C_{\a (\g} C_{\d ) \b} {\Bar {\cal 
F}} ~~~, 
\eqno(5.4) $$
where ${\Bar D}_{\ad} {\cal F} = 0$. The remaining non-vanishing
field strength superfields take the forms
$$ \eqalign{
F_{\a \, \un b \, \un c \, \un d} &=~ - \e_{\un a \, \un b \, \un c 
\, \un d} \, {\Bar D}^{\ad}\, {\Bar {\cal F}}  ~~~, \cr
F_{\un a \, \un b \, \un c \, \un d} &=~ i \e_{\un a \, \un b \, 
\un c \, \un d} \, \Big[ \, {D}^2 \,{\cal F} ~-~ {\Bar D}{}^2 \,{\Bar 
{\cal F}} ~ \Big] ~~~. }\eqno(5.5)  $$
As was given in the first considerations of irreducible super 
$p$-forms \cite{PFM}, the super 4--form defined by (5.4) and (5.5) is 
super-closed $((d {F})_{\un A \, \un B \, \un C \, \un D \, \un E} =
0$). 

The definition of the anomaly in terms of the $4$--forms (5.5) 
is
$$
S_{\rm BGJ} ~\equiv~ \frac{1}{4!} \, \int_{R^4} \, e^{\un a} 
\, e^{\un b} \, e^{\un c} \, e^{\un d} \, F_{\un a \un b \un c 
\un d} ~=~ \frac{i}{4} \, \int \, d^4x \, [ \, D^2 {\cal F} ~-~ 
\bar{D}^2 \bar{\cal F} \, ]
\eqno(5.6) 
$$ 
The essential problem is then reduced to one of specifying the form of 
the chiral superfield ${\cal F}$ in terms of a gauge parameter chiral 
superfield $\L$ and the YM gauge superfield $V$.

As in the bosonic case, one could in principle use the conditions 
in (5.2) and (5.3) to determine algebraically the supersymmetric 
BGJ anomaly. This would amount to consider the most general non-trivial 
linear combination of ``monomials'', expressed in terms of $e^V$, 
superconnections, field strengths and possibly their derivatives, 
linear in $\L$, and impose the conditions (5.2) and (5.3) 
to determine the coefficients of the linear combination.  Here we 
prefer to determine an explicit expression for the supersymmetric
anomaly by the use of a special homotopy operator along the lines of 
the work by McArthur and Osborn (MAO) \cite{MAO} who gave the
clearest and most succinct discussion of the issues involved with
the solution of the supersymmetric Wess-Zumino consistency conditions. 

In this reference, the anomaly is written as the sum of two terms 
$L$ and $\int_0^1 dyX(y)$ where $L$ is the covariant anomaly obtained 
from a regularized form of the one-loop effective action, and $X$ 
is a local functional added in order to satisfy the WZ consistency 
conditions.  It is constructed by using a homotopy, described by the
parameter $y$, which  is a class of maps denoted by $g_y$ satisfying 
the boundary conditions $g_{y = 0} =  {\bf I}$ and $g_{y = 1} = e^V$ 
(for our purpose, in a $K$-gauge where ${\Bar{g}}=1$).  In the class 
of maps $g_y$ satisfying the previous boundary conditions we choose 
the {\em minimal homotopy} defined as
$$
g_y  ~\equiv~ {\bf I} ~+~ y \, (~ e^V ~-~ {\bf I} ~) ~~~.
\eqno(5.7) $$
Its main merit is that it is {\it{linear}} in $e^V$.  Let us 
denote its inverse by ${\cal G}$ 
$$
{\cal G} ~\equiv~ g^{-1}_y ~=~ { {\bf I} \over {~~{\bf I} ~+~ y \, 
(~ e^V ~-~ {\bf I} ~) ~~ }} ~~\to~~ {\cal G}|_{y = 0} ~=~ {\bf I} 
~~,~~{\cal G}|_{y = 1} ~=~ e^{- V} ~~~,
\eqno(5.8) $$
which satisfies
$$
y \, {\cal G} \, e^V ~=~ {\bf I} ~-~ (\, 1 \, - \, y \,) \, {\cal G} ~~,
~~ y \, (\, e^V \, - \, 1\, ) \, {\cal G} ~=~ {\bf I} ~-~ {\cal G}
~~.
\eqno(5.9) $$
$$
{{\pa^{n} {~\,}} \over {{\pa y}^n }} \, {\cal G} ~=~ (-1)^n \, n! \, 
 (\, e^V \, - \, 1 \, )^n \, {\cal G}^{n + 1} ~~~~.
\eqno(5.10) $$

The minimal homotopy is uniquely characterized by the simplicity that 
is evident in the equations 
$$ \eqalign{ 
~\d_G g_y &=~ - i \, y \, (~ e^V \, \L ~-~ {\Bar \L } \, e^V  ~) ~~~,~~~ 
\d_G {\cal G} ~=~ i \, y \, {\cal G} \, (~  e^V \, {\L}\, 
-\, {\Bar \L} \, e^V   ~)  \, {\cal G} ~~~, \cr 
d  g_y &=~ - i \, y \, e^V \, [ ~ d \o^{\a} \, \G_{\a} ~
-~ d \o^{\Dot \a} \, {\Tilde \G}{}_{\Dot \a} ~+~ d \o^{\un 
a} \, (\, \G_{\un a} \,-\, {\Tilde \G}{}_{\un a} \,) ~]  ~+~ 
d y \, (~ e^V ~-~ {\bf I} ~)  ~~~, \cr
d {\cal G} &=~ i \, y \, e^V \, {\cal G} \,[ ~ d \o^{\a} \, 
\G_{\a} ~-~ d \o^{\Dot \a} \, {\Tilde \G}{}_{\Dot \a} ~+~ d 
\o^{\un a} \, (\, \G_{\un a} \,-\, {\Tilde \G}{}_{\un a} \,
)~] \, {\cal G} \cr 
&{~~~~}-~ d y \, (~ e^V ~-~ 1 ~) \, {\cal G}^2 ~~~, 
 } \eqno(5.11) $$
where $( d \o^{\a} , \, d \o^{\Dot \a} , \,d \o^{\un a} )$ are the basis
one-forms canonically dual to $( D_{\a} , \, {\Bar D}{}_{\Dot \a} ,
\pa_{\un a} )$.  Due to the form of (5.7), the connections in (5.11) 
are independent of the $y$-variable.  The second line in (5.11) allows 
the definition of a homotopically extended superconnection form via 
${\Hat \g} = i{\cal G} d g_y$.

The gauge superfield appears solely via exponential dependence.  This 
is the hallmark of the class of homotopy operators of the form $g_y =
{\bf I} + f(y) ( e^V - {\bf I})$. For the minimal one in (5.7), simple 
and explicit calculations show
$$ \eqalign{ 
{~~\,} [\, \d_G(\L^1) \,,\, \d_G(\L^2) \, ] \, 
\left(\begin{array}{c}  g_y \\
{\cal G} \\\
\end{array}\right)
&=~ \d_G(\L^3)  \left(\begin{array}{c}
g_y \\
{\cal G} \\
\end{array}\right) ~~~,~~~
\L^3 ~\equiv~ 
- i \, [\, \L^1 ~,~ \L^2 \, ] ~~~,  \cr
(~ d \, \d_G ~-~ \d_G \, d ~) \, g_y & =~ -~ [\Delta_1 + \Bar{\Delta
}_1] \, d  g_y ~~~, \cr]
(~ d \, \d_G ~-~ \d_G \, d ~) \, {\cal G} & =~-~ {\cal G} \left( [ 
\Delta_1 + \Bar{\Delta}_1]~d g_y \right) {\cal G} ~~~.} 
 \eqno(5.12) 
$$

We can now state the relation of the minimal homotopy to the irreducible
super 4--form field strength ${\cal F}$ in (5.5).   The minimal homotopy 
allows explicit evaluation of all quantities defined in ref. \cite{MAO}. 
The ``X''-function in our conventions is given by 
$$
\eqalign{ {~}
X(f_1,f_2) ~=~
\frac{1}{3} \int d^8 Z \, {\rm Tr}_s \Big[ ~ f_1 \cdot ({\Bar D}{
}^{\Dot \a}  f_2) \cdot {\Tilde {\cal W}}{}_{\Dot \a} \,+\, f_2 \cdot 
[ {\cal G} D^{\a} (g_y f_1 {\cal G}) g_y] \cdot {\cal W}_{\a} ~\Big]  
\,+\, {\rm {h.\,c.}} ~~~ , }
\eqno(5.13)
$$
where the functions $f_1$ and $f_2$ depend (again in our conventions) 
on the choice of the homotopy according to the following definitions
$$
f_1 ~=~ g_y^{-1} \, ({ {\pa  g_y }\over {{\pa y} }}) \, dy \qquad ,
\quad f_2 ~=~ {\cal G} \, \delta_R^1  g_y ~~~.
\eqno(5.14)
$$
The ${\rm Tr}_s$ operation is defined by
$$ \eqalign{
{\rm Tr}_s [\, {\cal A} \cdot {\cal B} \cdot {\cal C} \,] &=~ {\rm 
Tr} [\, \, {\cal A} ( {\cal B} {\cal C} \,+\, {\cal C} {\cal B} ) 
~+~ {\cal B} ( {\cal C} {\cal A} \,+\, {\cal A} {\cal C} )
~+~ {\cal C} ( {\cal A} {\cal B} \,+\, {\cal B} {\cal A} 
) \,\,] ~~~, \cr
{\rm Tr}_s [\, {\cal A} \cdot {\cal B}^{\a} \cdot {\cal C}_{\a} \,] &=~ 
{\rm Tr} [\, \, {\cal A} ( {\cal B}^{\a} {\cal C}_{\a} \,-\, 
{\cal C}_{\a} {\cal B}^{\a}  ) ~+~ {\cal B}^{\a} ( {\cal C}_{
\a} {\cal A} \,+\, {\cal A} {\cal C}_{\a} )
\cr 
&{~~~~~~~~~} ~-~ {\cal C}_{\a} ( {\cal A} {\cal B}^{\a} \,+\, 
{\cal B}^{\a}{\cal A}  ) \,\,]  ~~~.
} \eqno(5.15)$$

In the case of minimal homotopy one obtains 
$$
\eqalign{
f_1 &\equiv~ {\cal G} (e^V -1) ~~~,~~~ f_2 ~\equiv~ -iy {\cal G} \, 
e^V  \L ~~~, \cr 
{~~~~~~~}{\cal G}[ D_{\a} ( g_y f_1 {\cal G}) ]g_y \, & \equiv~ -ie^V
{\cal G}^2 \G_{\a} ~~~ ,~~~ {\Bar {D}}_{\dot \a}f_2 \, ~\equiv ~ 
y (1-y) \, e^V {\cal G} \Tilde{\G}_{\dot \a}{\cal G} \L ~~~,}
\eqno(5.16)$$
whereas the homotopically extended field strength ${\cal W }{}_{\a}$ 
and its ``tilde'' conjugate ${\Tilde {\cal W}}{}_{\Dot \a}$ are defined 
as follows,
$$ \eqalign{ {~~~~~~~~~~~~~}
{\cal W}_{\a} &\equiv~  y \, e^V {\cal G} \, w_{\a} ~=~ \Big\{ \, {\bf 
I} ~-~ (\, 1 \, - \, y\,) {\cal G} \, \Big\}  \, w_{\a} ~~~, \cr
w_{\a} &\equiv ~ W_{\a} ~-~ ( \, 1 \,- \, y \,) \, [ ~ {\Tilde \G}{}^{
\Dot \a} \, {\cal G} \, {\G}_{\un a} \,+\, ( \, 1 \,- \, y \,) \, {
\Tilde \G}{}^{\Dot \a} \,{\cal  G}\, {\Tilde \G}{}_{\Dot \a} 
\, {\cal G} \, {\G}{}_{\a} \cr  
& {~~~~~~~~~~} -\, i \, \fracm 12(\, {\Bar D}{}^{\Dot \a} {\Tilde
\G}{}_{\Dot \a}  ~-~ i \, {\Tilde \G}{}^{\Dot \a}\, {\Tilde \G}{}_{
\Dot \a} \, )\, {\cal G} \, \G_{\a} ~ ] ~~~, \cr 
{\Tilde {\cal W}}{}_{\Dot \a} & \equiv~  y \, e^V {\cal G} \, {\Tilde
w}_{\Dot \a} ~=~ \Big\{ \,  {\bf I} ~-~ (\, 1 \, - \, y\,) {\cal G} \,
\Big\}  \, {\Tilde w}{}_{\Dot \a}  ~~~, \cr   
{\Tilde w}{}_{\Dot \a} & \equiv ~ {\Tilde W}_{\Dot \a} ~+~  ( \, 1 \,
- \, y \,) \, [ ~ {\Tilde {\G}}{}_{\un a} \, {\cal G} \, \G{}^{\a} \,-
\, ( \,  1 \,- \, y \,) \, {\Tilde \G}{}_{\Dot \a} \,  {\cal G}\,
{\G}{}^{\a} \,  {\cal G} \, {\G}{}_{\a} \cr 
& {~~~~~~~~~~} -\, i \, \fracm 12 \, {\Tilde \G}{}_{\Dot \a} \, 
{\cal G} \, (\, {D}{}^{\a} {\G}{}_{\a} ~+~ i \, {\G}{}^{\a}\, {
\G}{}_{\a} )\,  ~ ] ~~~.}
\eqno(5.17) 
$$
Above, $W_{\a}$ is the standard field strength defined in (3.2), while 
the tilde quantities\footnote{Care should be taken to note that $W_{\a}$ 
and its ``tilde-conjugate'' are defined by (3.2) and (3.11), ${~~~~~}$ 
respectively. On the other hand, ${\cal W}{}_{\a}$ and its
``tilde-conjugate'' are defined by (5.17).} appearing on the RHS of eq.
(5.17) are defined in (3.11). 

We also work in chiral representation (with holomorphy manifest) to find
$$
\eqalign{ {~} 
&{\cal F} ~=~ - \frac{1}{2\pi^2} \,  \Bar{D}{}^2 \,\Big\{ \,  \, 
{\rm Tr} \, ( \,\Lambda \, \G^{\a} \, W_{\a} \, ) \, -\, \fracm 13 
\int_0^1 dy \, y \, {\rm  Tr}_s  \, \Big(\, e^V \, {\cal G} \, \L 
\cdot \, e^V \, {\cal G}{}^2 \, {\G}^{\a} \cdot {\cal W}_{\a} \cr
&{~~~~~~~~~~~~~~~~~~~~~~~~~~~~~~~~~~~~~~~\,~~~~~~~~~~}
\,+ \,(\,{\rm I} \, - \, e^V \,{\cal G} \,) \cdot  
e^V \, {\cal G}  {\Tilde {\G}}{}^{\Dot \a} \, {\cal G} \, \L
\cdot \Tilde{\cal W}{}_{\Dot \a} \,\, \Big) \, \Big\} \cr 
&{~~~~}\equiv ~ \Bar{D}^2 \, {\cal P}( \L\,; e^V){~~} ~~~.}
\eqno(5.18)  $$
Written in this form, only the inverse minimal homotopy ${\cal G}$
appears.  Making the symmetrization of the trace explicit we eventually
find
$$
\eqalign{  {~~~}
{\cal P} ~=~ - \fracm{1\,}{2\pi^2}& \,  \Big\{ \, {\rm Tr} \Big[ 
\, \L \, \Big( \,  \G^{\a} W_{\a} \,- \, \int_0^1 \, dy \, y ~ 
( \, 2 \, \Tilde{\cal W}^{\dot \a} \, \Tilde{\p}_{\dot \a} ~+~ 
\, [ \, {\cal W}^{\a} \, , \, \p_{\a} \, ] \, {\cal G} \, e^V \cr
& {~~~~~~~~~~~~~~~~~~~~~~~~~~~~~~~~~~~~~} ~-~ \{ \, \Tilde{\cal 
W}^{\dot \a} \, , \, {\cal G} \, e^V \, \} \, \Tilde{\p}_{\dot \a} 
~)~ \Big)~ \Big] ~ \Big\}  ~~~, } \eqno(5.19) 
$$
where we have defined
$$
\p_{\a} ~\equiv~  e^V \, {\cal G}^2 \, \G_{\a} \qquad , \qquad 
\Tilde{\p}_{\dot \a} ~\equiv~ {\cal G} \, (-\p_{\a})^{\dag} \, g_y
~=~ e^V \, {\cal G} \, \tilde{\G}_{\dot \a} \, {\cal G} ~~~.
\eqno(5.20)
$$
Therefore, upon defining $ d^6 Z \, \equiv \, d^4 x d^2 \q $ 
we find
$$
{\Tilde S}_{BGJ} (\L) ~\equiv~ i\fracm{1}{4}  \int d^6 Z ~ {\cal F}
~=~ i\fracm{1}{4}  \int d^8 Z ~ {\cal P}
\eqno(5.21) 
$$
for the holomorphic BGJ anomaly action.  

The superfield action given by (5.1) and (5.21) contains the component
action defined by (2.6) and (2.7).  Furthermore, using standard arguments,
the WZNW term can be obtained from the replacement
$$
{\cal P}(\L ; e^V) ~\to ~
\int_0^1 \, d w ~ {\cal P}( \L \, ;  {\cal U}{}^{\dagger} e^{ V } 
{\cal U} )  ~~~,~~~ {\cal U} ~\equiv~ e^{ - i  w {\L} } ~~~.
\eqno(5.22)
$$
(details concerning these results will be reported in an extended version
of this paper \cite{GGP}).   

One reason for the comparative simplicity of our result contrasted 
with those in ref.'s \cite{C22} and \cite{MAO}, is precisely the use
of the minimal homotopy.  A non-minimal choice of the homotopy that
has been widely discussed previously is defined by ${\tilde g}{
}_y \equiv \exp{[\,y ~V \,]}$. The gauge variation of this expression 
is vastly more complicated than the first result given in (5.11).  
Many other previous expressions for the anomaly are non-minimal (i.e.
contain cohomologically trivial terms) as can be seen explicitly in many
places (e.g. the work by Guadagnini, Konishi and Minchev \cite{C22}). 
Cohomological and topological non-minimality appear to be the source of
much of the opacity of the literature on the topic of supersymmetric BGJ
anomalies. \newline
${~~~}$ \newline
${~~~~~~~~~}$``{\it {The labour we delight in physics pain.}}'' -- W.
Shakespeare
$${~~~}$$
\noindent
%%%%%%%%%%%%%%%%%%%%%%%%%%%%%%%%%%%%%%%%%%%%%%%%%%%%%%%%%%%%%
{\bf {Acknowledgment}} \newline \noindent
%%%%%%%%%%%%%%%%%%%%%%%%%%%%%%%%%%%%%%%%%%%%%%%%%%%%%%%%%%%%%
${~~~~}$The authors wish to acknowledge discussions with 
I. McArthur and W. Siegel. S.P. thanks the Physics Department of 
University of Maryland for hospitality during the period when
part of this work was performed. S.J.G. likewise acknowledges 
the hospitality of Dipartimento di Fisica dell'Universit\'a 
di Milano--Bicocca and INFN, Sezione di Milano.   

\newpage

%%%%%%%%%%%%%%%%%%%%%%%%%%%%%%%%%%%%%%%%%%%%%%%%%%%%%%%%%%%%%%%%%%%

\end{document}